# The effect of longitudinal debonding on stress redistributions around fiber breaks: Incorporating fiber diameter distribution and fiber misalignment

M. Jafarypouria[*], S.V. Lomov, S.G. Abaimov

Center for Petroleum Science and Engineering, *Skolkovo Institute of Science and Technology*, *Moscow*, *Russia*

[*]Corresponding author: Milad.Jafarypouria@skoltech.ru

## Abstract

This research explores the influence of interfacial debonding between a broken fiber and matrix on stress redistribution surrounding a fiber break within a unidirectional (UD) impregnated fiber bundle, accounting for misalignment of fibers and fiber diameter distribution in randomly packed fiber configurations. Finite-element modelling is conducted on carbon-reinforced epoxy UD bundles with one fiber broken for different combinations of the bundle parameters: aligned/misaligned fibers and constant/randomly distributed fiber diameters. Two definitions of stress concentration factor (SCF) are examined, based on average and maximum stress over the fiber cross-section. The study reveals a statistically significant difference for average SCF for misaligned bundles with both constant and variable diameters of fibers compared to the case of aligned bundles of fibers with constant diameter. When the calculated SCFs are incorporated in a bundle strength model, the failure strain of unidirectional composites can be more realistically predicted.

*List of symbols*

| | | | |
|---|---|---|---|
| AFCD | Aligned fiber with constant diameter | max | maximum |
| AFDD | Aligned fiber with diameter distribution | $E$ | Young's modulus |
| MFCD | Misaligned fiber with constant diameter | $G$ | Shear modulus |
| MFDD | Misaligned fiber with diameter distribution | $v$ | Poisson's ratio |
| SCF | Stress concentration factor | $\sigma_y$ | Yield stress |
| IL | Ineffective length | $\sigma_u$ | Maximum plastic stress |
| $l_d$ | Debonded length | $L$ | RVE length |
| IFBM | Impregnated fiber bundle model | $\emptyset$ | RVE diameter |
| $\sigma$ | Experimentally measured standard deviation of the fiber diameter distribution | $N_E$ | Total number of elements in RVE |
| VF | Fiber volume fraction | $\sigma_{z,avg}$ | average fiber stress |
| *maxSCF* | Maximum SCF | RAFCD | Regression in the AFCD case |
| $D_{mean}$ | Mean fiber diameter | $d/R$ | Relative distance to the broken fiber, with $R$ being fiber radius |
| $D_i$ | Minimum fiber diameter | $\bar{\mu}$ | Mean |
| $D_o$ | Maximum fiber diameter | $p_{value}$ | Significant statistical difference |
| CV | Coefficient of variation | BF | Broken fiber |
| RVE | Representative volume element | NNFs | Nearest neighbor fibers |
| $D$ | Fiber diameter | CZM | Cohesive zone model |
| Avg | Average | EE | Embedded element |
| Std | Standard deviation | C3D8R | Linear, hex-dominated elements with reduced integration |
| min | minimum | C3D20R | Quadratic, hex-dominated elements with reduced integration |
| $N_F$ | Number of fibers | | |

## 1. Introduction

Fiber-reinforced polymer composites show different failure modes; however, the final failure of a composite at tensile loading is typically caused by the longitudinal tensile failure of 0° UD ply in a laminate or 0°-oriented yarn of a textile reinforcement. The main damage mechanisms of an axially loaded UD composite include fiber breakage and fiber-matrix interfacial debonding.

The modelling efforts in fiber breaks during longitudinal tensile failure of UD composites started in the 1960's, 1970's, and 1980's [1-5], and became a fast-developing topic in recent years [6-16]. All models are essentially based on the same two principles; (I) the fiber strength scatter follows a Weibull distribution, and (II) when a fiber breaks, it locally loses its longitudinal load transfer capability, which leads to stress concentrations on the surrounding fibers. The readers are referred to in-depth and comprehensive publications [6,17-21] and dissertations [22,23] in this area. In the following, we outline the essential parameter definitions used in the current study.

*Stress redistribution around a fiber break*

Near a fiber break, the redistributed stress results in a relative stress enhancement in the intact nearest neighbor fibers (NNF) as compared to the nominal stress, which is commonly quantified by stress overload factor, commonly called stress concentration factor (SCF). Even after the failure, the broken fiber continues to carry the load at locations further from the point of failure. According to Rosen [1], the length of significant stress reduction is called ineffective length (IL) and is defined as twice the fiber length over which 90% of stress recovery occurs. More details on SCF and IL are given in section 2.4.

The radial gradient of axial stress is highly pronounced, with stress rapidly diminishing from a fractured fiber in the radial direction. This effect is anticipated to be particularly significant when examining fibers of varying diameters. The baseline models [18-20] employ the average SCF taken as the stress averaged over the cross sections in the neighboring undamaged fibers. An alternative approach is to use the maximum stress in a cross-section as a criterion for the fiber break [23-25], particularly because the strong stress localization is supported by experimental evidence [26]. The reason behind uneven stress distribution at NNFs cross-sections is related not only to the rapid decay of axial stress in radial direction. In Fig.SUPP1, one can observe at NNFs cross-sections the uneven axial stress distribution. At the side closer to the break, the stress is higher than the axial stress in fibers far from the break point. At the side of the cross-section most distant from the break point, the stress is lower than the axial stress in fibers far from the break point. This signifies strong bending of the NNFs in the vicinity of the break point leading to enhancement of stress localization over cross-sections. Our current investigation seeks to determine and compare the stress concentration factor (SCF) values using both formats (average and maximum over NNF's cross-section). Section SUPP 1 in Supplementary materials presents details of the local stress distribution over a fiber cross-section.

*Interfacial debonding*

A crucial hypothesis relates to the stress singularity surrounding a fiber break. For elastic, well-bonded materials, stress concentrations in the matrix surrounding a fiber break are infinite. Earlier models assume perfect bonding and an undamaged matrix [27,28]; introduction of interfacial debonding, matrix plasticity, and matrix cracking in the fiber break plane is expected to resolve the stress singularity at the fiber break.

Fiber-matrix debonding usually happens for weak interfacial bonds and refers to the separation of the fibers from the polymer matrix. Debonding causes an increase in the ineffective length [29,30], but decreases the SCFs on the intact fibers [30-33]. Fiber-matrix debonding have been extensively investigated [34-36].

AhmadvashAghbash et al. [37,38] constructed a comprehensive finite-element models to investigate longitudinal debonding between fiber and matrix associated with fiber break, which account for all contributors to the longitudinal debonding (interfacial strength, fracture toughness, thermal residual stresses, friction, and matrix plasticity) in regular and randomly packed fiber configurations. The findings demonstrated a significant decrease in the stress concentration factor (SCF) in NNFs within a bundle of constant (identical) fiber diameters (FCD) when fiber/matrix debonding occurs. Zhuang et al. [39] utilized an axisymmetric FE model to explore the progression of failure starting from a central fiber break surrounded by intact fibers in two scenarios: I) a matrix crack initiates from the fractured fiber end and propagates normal to the fiber axis, and II) a matrix crack deviates from a short fiber/matrix debond crack and extends towards neighboring fibers. The highest SCF and probability of failure in NNFs occur when the matrix crack front reaches the fibers, with the extent of debonded length of the broken fiber influencing both the SCF and the axial length over which the SCF is present. Their results also show that the inclusion of interfacial debonding significantly decreases the SCF compared to models with well-bonded interfaces.

The extent of debonding near a fiber break is affected by interfacial properties, particularly for Mode II fracture, which are variable for various composite systems and show significant scatter depending on the method of characterization [26,40-42]. The progression of debonding is highly sensitive to the local fiber configuration [43,44]. The present study aims at revealing the effect of the presence of debonding on the SCF sensitivity to modeling parameters such as fiber misalignment and fiber diameter distribution.

*Matrix plasticity and thermal residual stress*

The epoxy resins are typically modeled as elastic or visco-elastic materials. This presumption, which is based on observations of bulk specimens, is not indicative of how the epoxy matrix behaves in high-performance fiber-reinforced composites with small interfiber volumes. Hobbiebrunken et al. [45] reported a high degree of plasticity in small-scale epoxy specimens. Plasticity affects the typical stress profiles in both intact and broken fibers directly because it distributes the load over a larger volume. In contrast to elastic deformation, the maximum SCFs of the neighbouring intact fibers decrease in presence of plasticity.

The mismatch in thermal expansion between a matrix and fiber causes thermal residual stresses. The most severe mismatch typically occurs between carbon fibers and polymers because of the longitudinal contraction that carbon fibers experience when heated. Before a fiber break happens, the arising stresses may already be causing shear between the fiber and matrix (fiber experiences compressive strain according to van den Heuvel et al. [31]). The mismatch in the radial coefficients of thermal expansion persuades residual stresses (normal forces), therefore enabling interfacial friction on the debonded surfaces. The simple analytical approaches fail to consider the stress state complexity in the vicinity of a fiber break or more complex loading through thermal residual stresses. These inadequacies were remedied by 2D [46] and 3D [28,30] FEMs.

The effects of matrix yielding and thermal residual stresses have been demonstrated in the SCF values, IL, and $L_d$ lengths reported by AhmadvashAghbash et al. [38]. The finite-element models utilized in this study incorporate matrix plasticity and thermal residual stresses as well.

*Fiber misalignment*

The misalignment of fibers in IFBMs is an important factor that may affect the location and extent of fiber breaks. Although the link between misalignment and fiber breaks is considered weak [23], further research is needed to fully understand this relationship. When fibers are not aligned properly with the direction of the applied load, stress concentrations may develop at specific points along fiber's length due to misalignment, potentially increasing the probability of fiber failure. Experimental methods like microscopy, X-ray imaging, and digital image correlation have been utilized in various studies to investigate fiber misalignment and its potential association with fiber breaks [48-54].

Fiber misalignment, a common defect in fiber-based materials, has a significant effect on the mechanical properties and performance of unidirectional (UD) composites, affecting factors such as tensile strength and modulus [55-57]. The sources of fiber waviness vary depending on the composite manufacturing process and processing parameters [58-61]. In a UD ply, deviations from the intended fiber direction (misalignment) can occur both in-plane and out-of-plane.

Misalignment of fibers, a critical element influencing compressive strength [62-65], is often overlooked in models for longitudinal tension. Current fiber-break models assume straight and parallel fibers, which do not represent real materials accurately [66,67]. Jafarypouria et al. [68] introduced the concept of misalignment within fiber bundles as a long-wave defect. The study revealed that misaligned fiber significantly affects the SCFs, more pronounced in NNFs, where the effect of fiber misalignment is more localized.

*Fiber diameter distribution*

Fiber diameter distribution refers to the range of diameters of individual fibers present within a composite material. This distribution can vary depending on the manufacturing process and the type of fibers used. There are various instruments and methods for measuring fiber diameters. Mesquita et al. [69]

employed an automated single fiber tensile testing approach to generate extensive data sets [70] of individual carbon and glass fiber properties. The fiber diameter was assessed using a laser diffraction system. Different fiber diameters were incorporated for different types of fibers in hybrid composites reported in [13,16,71-73], but without introducing the diameter variability inside a fiber type.

The spatial position of fibers in a composite is influenced by various factors during the composite manufacturing process, such as applied pressure, resin flow, and the constraints of the mold, etc. However, even under optimal manufacturing conditions, fibers are typically positioned irregularly, leading to clustering and the development of resin-rich areas. Various statistical metrics are available to measure this irregularity, i.e. fiber position (spatial distribution), fiber cross-sectional shape (diameter distribution), and fiber orientation (alignment distribution) [74-76].

Using data sets from [69,70] as the foundation for our earlier research [77,78] on fiber diameter distribution (FDD) characterization, our finding illustrated that introduction of FDD leads to significant decreases in SCF for carbon fiber bundles compared to fiber constant diameter bundles at different VFs.

The formation of localized fiber break clusters depends on SCFs in the neighboring intact fibers, as well as the length over which the SCFs are active, which is characterized by debonding and ineffective lengths of the broken fiber. The aim of this research is to bridge the gap between numerical and experimental data by integrating key modeling assumptions to reliably forecast SCFs. This involves exploring the sensitivity of SCFs to significant factors like fiber-matrix debonding, fibers misalignment, and fiber diameter distribution.

## 2. Model description

### 2.1 Geometry

We utilize a modified version of the Catalanotti algorithm [79] to generate Representative Volume Elements (RVEs) containing distributions of fiber misalignment angles (both in-plane and out-of-plane) and fibers of varying diameters in a random manner.

A cylindrical 3D RVE with one of three nominal fiber volume fractions (VFs), VF30%, VF50%, and VF60%, is produced by truncating the generated square 2D RVE to a circular RVE and then extruding along its normal axis for AFCD and AFDD bundles, while for MFCD and MFDD bundles, it is extruding along a path specific for each fiber. In the state-of-the-art models, the RVE needs to be large enough to ensure that the ineffective length will be well captured: the RVE should be sufficiently long in the fiber longitudinal direction. RVEs with lengths ranging from $7.5 \times D$ to $20 \times D$ were used in several studies [14,16,80], where $D$ is the fiber diameter (7 μm for carbon fibers). This, however, is not sufficiently large for a debonding model, and the model's length ($L$) was established at $285 \times D$. This large dimension helps to reduce the apparent stiffness loss of the simulated object caused by a fiber break and debonding [30,37,38].

### 2.2 Bundles generation

In order to create a bundle with variable fiber diameters, using the data presented in Table 1, a normal distribution is employed, as investigated in our earlier study [77,78]. Results in [77,78] show that the influence of the diameter variability on SCF is weak (in case of average calculations); in order to demonstrate this influence in a clear way, the case of high variability is considered here, with standard deviation of the diameter distribution equal to $2\sigma$, where $\sigma$ is the experimentally measured distribution's standard deviation [69]. To generate a bundle with fiber misalignment, the methodology from the previous study is adopted [68]. This involves creating 3D RVEs with a spatially-correlated random distribution of fiber misalignment, following the measurements conducted in the literature. Table 1 characterizes the generated fiber misalignment angles of all bundle types. It shows the spread, mean, and standard deviation of the maximum local fiber misalignment for five bundles generated at different VFs.

## 2.3 Material, FE model, and mesh verification

The plastic response of the epoxy matrix is simulated using the ABAQUS plasticity model, prescribed with estimated true stress-strain values with epoxy properties presented in Table 2.

FE models of impregnated bundles, for example, in [19], use continuous meshes to create RVEs by dividing the volume into different domains, such as reinforcement and matrix, through partitioning or subtraction processes (refer to 'direct' models). Applying 2% strain to such a mesh composed of tetrahedral elements often leads to convergence issues due to their tendency to distort more easily. When modeling fiber misalignment, the necessity for these elements cannot be overlooked in order to generate an RVE that accounts for fiber misalignments, despite their inherent instability. A high-density mesh with second-order elements is typically necessary to effectively simulate matrix deformation using tetrahedral elements. This results in a substantial number of 3D-Tet elements, leading to significant computational effort in finite element modeling due to the abundance of higher-order elements.

Superimposed meshes, or embedded elements (EE), overcome the highlighted drawbacks of direct FE modelling. This technique has been investigated for analysis of various composite materials and structures [81-84]. Embedded elements can be combined with cohesive zones modelling (CZM), which are a common method to simulate interfacial damage, as proposed in [85,86]. Moreover, embedding can be done mechanically [87-90] or thermally [91].

All fibers apart from the central (broken) fiber, are meshed independently of the matrix, and then the fibers are embedded into the matrix using "constraint embedded region" in ABAQUS (see Fig 1). The mesh in fibers and matrix uses linear, hex-dominated elements with reduced integration (C3D8R). The decision to use this element type is supported by a comparative analysis conducted by Breite [23], which demonstrates that there are no distinguishable differences in results when using either the quadratic (C3D20R) elements or the linear element type (C3D8R). In the present study, the analyses are performed in ABAQUS/Standard. The fibers and matrix are generally defined as embedded and host domains, respectively. The nodal displacements of the embedded domain are interpolated from those of the host domain so that the two domains are coupled together.

The CZ is applied only on the broken central fiber, which is not embedded, but meshed continuously with the matrix (see Fig 1c). The neighboring fibers do not develop debonding during stress redistribution, therefore CZ are not introduced on them. This approach reduces the computational complexity.

The broken fiber-matrix interface is modelled by a surface-based cohesive contact with a bilinear traction-separation law. The corresponding FE simulation follows the model of Jafarypouria et al. [68,77]. For more details about the FE models and material used see Table 1 and Table 2.

The matrix in EE approach occupies full volume of the model, including the volume ''under'' the reinforcement (except broken fiber). This creates volume redundancy [83]. To resolve the volume redundancy in the embedded elements, different approaches have been applied [82,83]. In all these approaches, the presented analysis is restricted to continuous stress–strain fields, prior to damage. An open question is: Does "neglecting volume redundancy" lead to any numerical artefacts, presumably close to the break plane, which would prevent using this idea for prediction of stress redistributions? Very limited attempts have been accomplished so far to introduce interfacial damage into the framework of the EE models using 3D FE [89]. The EE technique is used in interfacial damage models, while disregarding volume redundancy. The related errors are negligible in the present case, as test calculations show negligible difference between the direct meshing (no volume redundancy) and EE model at calculations of $maxSCF_{avg}$ and $maxSCF_{max}$ for carbon bundles at VF50%. The reader is referred to Supplementary materials, Fig. S3, for this comparison.

Fig. 1d shows the applied boundary conditions. The entire black plane of RVE is displaced with a displacement of 0.02×$L$ ($L$ is the RVE length) corresponding to a maximum nominal strain of 2% (uniaxial tension). Loading was implemented through applying displacements on corresponding face (matrix, broken fiber, and EE nodes). The lateral surface of the RVE is set as traction-free. X-symmetry boundary condition is applied to the RVE at the plane of the fiber break plane, except fiber's cross-sectional area and its perimeter (highlighted in gray) to simulate the break. The planar matrix crack around the fractured fiber can be incorporated into the model by substituting the symmetry condition with a traction-free boundary on the corresponding circular matrix crack area. Because of introducing the interface cohesive elements [37,38], the release of the nodes at the perimeter of the broken fiber (as in Ref. [9]) is redundant.

Fig. 2 illustrates an example of axial stress redistribution in a bundle for a misalignment random realization at VF50%: Fig 2a shows the stress recovery profile of the broken fiber; Fig 2b depicts the SCF profile on the nearest intact fiber in the bundle; Fig 2c represents the axial stress in BF and its NNFs; and Fig 2d shows the damage distribution at the debond front on the surface of the broken fiber at 2% strain.

The validity of the modelling should check the dependence on the size of fiber and matrix meshes. For the transversal direction, the matrix approximate global mesh size is chosen to be 2 (0.57 times the fiber radius ($R$), where model units are micrometers). Accordingly, the matrix mesh around the break region ranges from the smallest approximately $0.08 \times R$ to the largest matrix mesh about $0.72 \times R$ far from the break region. The smallest cross-sectional fiber mesh size in the BF and its NNFs is approximately $0.07 \times R$. The smallest longitudinal size of fiber and matrix meshes around the break region is approximately $0.28 \times R$ to the largest mesh far from the break region, which is about $70 \times R$. In accordance with a mesh sensitivity analysis, a finer mesh is applied to the broken fiber, the surrounding matrix, and the NNFs in the proximity of the plane of the broken fiber. For computational efficiency, the mesh density is gradually reduced as it extends radially and axially from the fiber breakage and its plane (refer to Fig. 1a,b and c).

In a mesh-sensitivity study [38], the direct method with CZM between broken fiber and matrix was analyzed using C3D20R elements. The smallest element size of 197 nm × 35 nm × 940 nm was applied to the broken fiber, surrounding matrix, and neighboring fibers facing the break. The current study employs C3D8R elements with an element size of 61 nm × 200 nm × 366 nm. In our study, the fiber has 100 elements on perimeter.

Section SUPP 2 in Supplementary materials presents a comparison study supporting the accuracy of the applied mesh and modeling approach in this subsection.

## 2.4 SCF calculations

Two definitions of stress concentration factor (SCF) are analyzed, considering the average and maximum stress over the fiber cross-section. For average SCF, Swolfs' definition [80] was used, which involves calculating the ratio of the relative increase in average fiber stress $\sigma_{z,avg}$ at $z^*$ divided by the average fiber stress $\sigma_{z,avg}$ far away from the failure plane:

$$SCF_{avg}(z=z^*) = \frac{\sigma_{z,avg}(z=z^*) - \sigma_{z,avg}(z=L)}{\sigma_{z,avg}(z=L)} \times 100 \qquad (1)$$

where $z^*$ is the distance along fibers from the break plane and $\sigma_{z,avg}$ defined as the longitudinal fiber stress of the elements in that plane averaged over fiber's cross-section. All SCF results will be expressed as percentages multiplied by 100. Fig 2b depicts variation of the defined SCF as a function of distance along the length of the intact fiber, which is the nearest neighbor of the broken fiber. The maximum of the average SCF along the fiber ($maxSCF_{avg}$) is achieved very close to the break plane, after which the stress quickly decays by moving away from the failure plane.

In addition to the average SCF, we also take into account the maximum SCF, which is determined by the maximum stress at the cross-section of the fiber. In this case, $\sigma_{z,avg}$ in Eq (1) is substituted by $\sigma_{z,max}$, representing the maximum value of axial stress in the fiber cross-section (Eq (2)). The maximum ($maxSCF_{max}$) is then identified along the length of the fiber. An illustration of the location of this peak and the stress distribution on the fiber cross-section can be seen in Fig. 2c, inset.

$$SCF_{max}(z = z^*) = \frac{\sigma_{z,max}(z = z^*) - \sigma_{z,max}(z = L)}{\sigma_{z,max}(z = L)} \times 100 \tag{2}$$

To assess the effect of debonding within a AFCD bundle on stress redistribution around a single fiber break and judge the statistical significance of the differences relative to debonding in AFDD, MFCD, and MFDD bundles, the following methodology was utilized. Let $SCF_r(d/R)$ be a regression of the reference SCF curve (AFCD bundle), $SCF_d(d/R)$ is the SCF value for fibers in a AFDD model, $SCF_m(d/R)$ is the SCF value for fibers in a MFCD model, and $SCF_{md}(d/R)$ is the SCF value for fibers in a MFDD model. Then statistical analysis is applied to differences $\Delta = SCF_d(d/R), SCF_m(d/R), SCF_{dm}(d/R) - SCF_r(d/R)$, analysing a hypothesis $\Delta \neq 0$. The comparison of the SCFs follows a one-tail t-test with an unequal variance analysis.

**Table 1.** Model input parameters used for RVE' generation: AFDD parameters of carbon T700s fibers, and maximum misalignment angles in random realizations of MFCD and MFDD models.

| | | FDD [77] | | | | | | |
|---|---|---|---|---|---|---|---|---|
| | $D_{mean}$ (μm) | | | | 7 | | | |
| | $D_i$ (μm) | | | | 6 | | | |
| | $D_o$ (μm) | | | | 7.8 | | | |
| | CV | | | | 4.6% | | | |
| | | MFCD | | | | MFDD | | |
| | | Min (°) | Max (°) | Avg | Std | Min (°) | Max (°) | Avg | Std |
| Nominal VF (%) | 30 | 2.1 | 4.3 | 3.0 | 0.7 | 2.0 | 4.2 | 3.1 | 0.7 |
| | 50 | 1.4 | 3.2 | 2.3 | 0.5 | 1.3 | 2.8 | 2.0 | 0.5 |
| | 60 | 1.3 | 3.0 | 2.2 | 0.5 | 1.3 | 1.9 | 1.7 | 0.2 |

**Table 2.** Model input parameters – constituents' properties: carbon fiber properties, epoxy properties with linear hardening, thermal expansion coefficients, and default carbon-epoxy interface properties.

| T700S carbon fiber [38] | |
|---|---|
| Fiber diameter, $D$ [μm] | 7 |
| $E_{11}$ [GPa] | 230 |
| $E_{22}$ [GPa] | 15 |
| $G_{12}$ [GPa] | 13.7 |
| $\nu_{12}$ [ – ] | 0.25 |
| $\nu_{23}$ [ – ] | 0.25 |
| Longitudinal CTE, $\alpha_{11}$ [$10^{-6}$/°C] | −0.38 [92] |
| Transverse CTE, $\alpha_{22}$ [$10^{-6}$/°C] | $10^a$ |
| Sicomin SR8500/KTA313 epoxy [23] | |
| Tensile modulus [MPa] | 3687 |
| Shear modulus [MPa] | 1073 |
| $\nu_m$ [ – ] | 0.405 |
| Yield stress [MPa] | 86 |
| Maximum plastic stress [MPa] | 110.0 |
| $\alpha_m$ [$10^{-6}$/°C] | 62.5 [93] |
| Default interface | |
| $\tau_c$ [MPa] | $40^a$ |
| $G_C$ [J /$m^2$] | $50^a$ |
| $\mu_i$ [ – ] | 0.3 |
| Interfacial stiffness, $K_i$ [MPa /mm] | $50^a$ |

a Selected within the range of values, reported in literature[42,43,96,97].

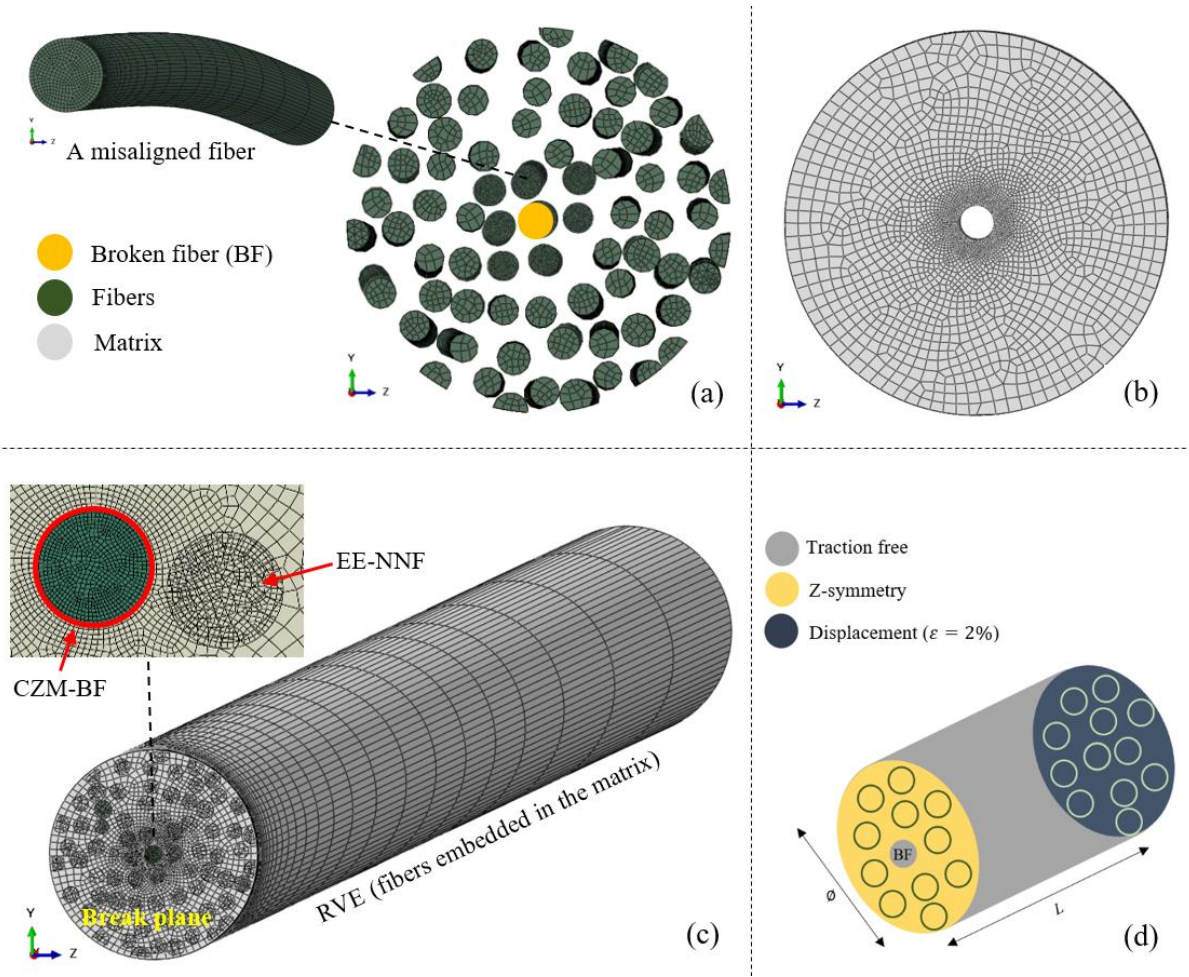

**Fig. 1.** Description of the model depicting the (a) transversal mesh density on the fibers, (b) transversal mesh density on the matrix (c) RVE with longitudinal mesh density, (d) boundary conditions and model dimensions.

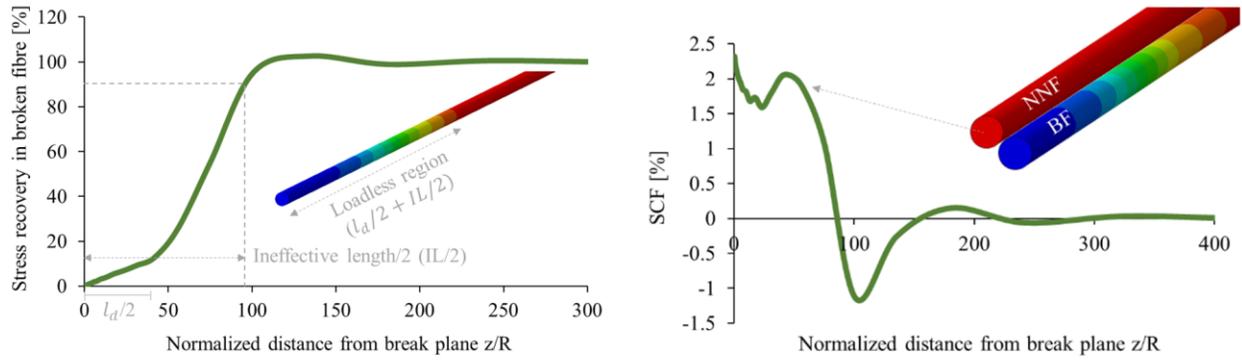
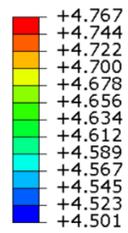
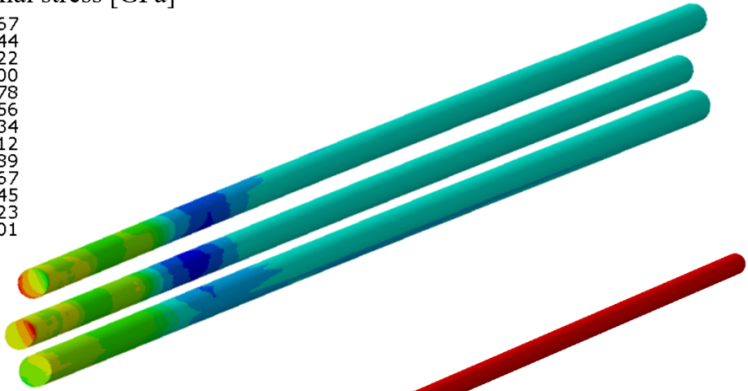
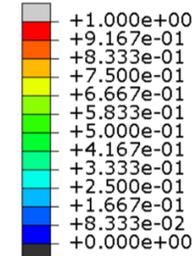
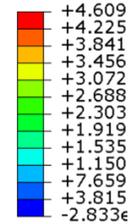
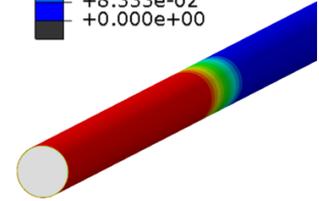

**Fig. 2.** Axial stress redistribution for a misalignment random realization with VF50%: (a) stress recovery profile of the broken fiber, (b) the SCF profile on the nearest intact fiber, (c) cutaway RVE with SCF on three intact neighboring fibers and axial stress on the broken fiber (the broken fiber is isolated from the RVE to facilitate the perception), and (d) damage distribution at the debond front on the surface of the broken fiber (for a better visualization only a small portion of the whole broken fiber is shown). Insets demonstrate axial stress distribution along the fibers.

## 3. Results and discussion

In this section, the stress redistribution (SCF and IL) for impregnated fiber bundle in AFCD, AFDD, MFCD, and MFDD models, for three nominal fiber volume fractions (VF) of 30%, 50%, and 60% are presented. The SCF and IL are investigated for debonded broken fiber (BF)/matrix interface. The SCF is calculated based on average normal stress ($maxSCF_{avg}$) and peak normal stress ($maxSCF_{max}$) over the cross section. To evaluate the effect of fiber-matrix debonding interface accounting for fiber misalignment and fiber diameter distribution on the stress redistribution parameters, the SCFs and ILs are calculated with the applied average strain of 2%, i.e. in the inelastic regime of deformation, which reveals in full effects of non-linearity of the matrix behavior.

### 3.1. Average normal stress over the cross section

Five random realizations were generated for each fiber volume fraction. The models use the material properties and FE parameters presented in Table 1 and Table 2.

### 3.1.1 SCF

Fig. 3 compares stress concentrations, Eq (1), for all four carbon bundle types (AFCD, AFDD, MFCD, and MFDD) at different VFs in the neighbor intact fibers around a single fiber break. The maximum average SCF ($maxSCF_{avg}$) is shown as a function of the relative distance ($d/R$) from the broken fiber. Fig. 3a,c,e illustrate SCF trends for AFDD, MFCD and MFDD models in comparison with AFCD models.

It is seen that the **AFDD** bundles have higher SCF than AFCD bundles spanning the entirety of *d/R* range at VF30%, but this becomes less pronounced at higher VFs. As the volume fraction of fibers increases, the overall stress distribution within the bundle becomes more uniform, leading to a reduction in the difference in SCF between the AFDD and AFCD bundles. This contrasts with our observations in perfectly bonded bundles and debonded bundles based on a fixed debonding [78] (the broken fiber has constant debond, with the size as the carbon fiber presented in AhmadvashAghbash et al. [38]; the debonded portion of the fiber near the break was freed). The observation from [78] suggests that in that specific scenario, the effect of the static fiber/matrix debonding on SCF in neighboring fibers is different compared to when the debond propagation happens. This could be due to factors such as interface properties, bonding between fibers and matrix, or other mechanical interactions within the bundle.

The **MFCD** bundles represent a significant lower SCF trends compared to AFCD bundles. Again, this differs from the findings of our earlier study involving perfectly bonded bundles [68]. In constant fiber diameter bundles without debonding, the misalignment of fibers leads to increased stress concentration at the interface between the fibers, resulting in a higher stress concentration factor. This is because the misaligned fibers create discontinuities and irregularities in the load transfer path, leading to localized stress concentrations. However, in the present study (debonded fiber bundles with misalignment), the presence of the debonding interface between the fiber and matrix can act as a stress relieving mechanism. When a load is applied, the debonded interface allows for some degree of slippage or movement between the fiber and matrix, which can help to distribute the load more evenly and reduce stress concentrations at the interface. This can result in a lower stress concentration factor compared to constant fiber diameter bundles without debonding.

The **MFDD** bundles show significantly lower SCF trends compared to the AFCD bundles, but display an increasing trend when compared to the MFCD bundles. This is due to the variability in fiber diameters within the bundles, which leads to an increase in SCF due to important reasons: (i) uneven load distribution; fibers with different diameters will possess varying load-bearing capacities. When subjected to external loads, the bigger fibers will bear more load compared to the smaller fibers (see an earlier study [77,78]). This uneven load distribution can lead to localized stress concentrations at the interfaces between different fiber diameters, (ii) interfacial bonding; the interface between fibers and matrix plays a crucial role in transferring loads from the matrix to the fibers. Variability in fiber diameters can affect the interfacial bonding quality. Bigger fibers may have weaker bonding with the matrix, leading to stress concentrations at these weak points. Section SUPP 3 in Supplementary materials presents an investigation supporting the effect of fiber diameter on $l_d$ and IL.

As it is seen in Fig.3a,c,e, all bundle types depict a considerable increase in the peak of the $maxSCF_{avg}$ graphs ($maxSCF_{avg}$ in NNFs) at different VFs. This demonstrates that the SCF is more influenced in NNFs. By increasing VF, due to the presence of fibers at smaller distances from the broken fiber, the difference becomes stronger in the peak of the $maxSCF_{avg}$ graphs compared to far away distances ($d/R \leq 2$). Moreover, the peak value of SCF progressively decreases with higher volume fraction.

In the earlier debonded randomly packed model of a AFCD bundle [38], the peak of the SCFs in NNFs was reported to be 5.3%, 2.9%, and 1.7% for VF30%, 50%, and 70%, respectively. In our study, for a similar carbon-epoxy system and applied strain 2%, the peak SCF of the NNFs in 30%, 50% and 60% VF cases was calculated to be, respectively, 3.6%, 2.9%, and 2.4% of the values in the MFDD bundle, which is a different prediction than [38] for the AFCD bundle. This reflects the influence of both fiber misalignment and variation in fiber diameters.

Fig. 3b,d,f represent the $maxSCF_{avg}$ difference between the AFDD, MFCD, and MFDD models and the corresponding regression in the AFCD case (RAFCD). The data points for AFDD exhibit a positive difference with RAFCD at different VFs, while in the case of MFCD and MFDD, they demonstrate a negative difference with RAFCD at VF30% and VF50% but a positive difference with RAFCD at VF60%. The statistical analysis of differences is presented in section 3. Due to the nature of variation of fiber diameter in AFDD bundles, the value of the normal stress in the nearby fibers varies significantly, resulting in higher scattering of SCF data points for the AFDD bundles in comparison with the AFCD models. This is more pronounced for MFDD bundles as the presence of misaligned fibers in MFDD bundles further intensifies this variability, leading to even greater scattering of SCF data points compared to the AFDD bundles. This is more essential at places close to the fiber break. An interesting finding is that the trends for MFCD and MFDD bundles shift from negative to positive when compared to the AFDD bundle. This indicates that the effect of fiber misalignment becomes more significant at higher VF. This phenomenon is highlighted in reference [68], which demonstrates that as the VF increases, the contrast between misaligned bundles and perfectly aligned bundles becomes more pronounced.

### 3.1.2 Debond length

Fig. 4 shows the average half debond and ineffective lengths for all cases at VF30%, 50%, and 60% with their corresponding standard deviations. Ineffective length is measured from the fiber break plane, referred to as a loadless region (see Fig. 2a). It is evident that $l_d$ and IL grow with an increase in VF. In fibrous polymer composites, increasing the fiber volume fraction can result in a longer debond length through debond propagation. With a higher fiber volume fraction, there are more interfaces between fibers and matrix in the bundle. These interfaces serve as potential locations for debonding to start and spread. A higher VF increases the likelihood of stress concentrations at these interfaces under applied loads. Consequently, this can lead to more extensive debonding and longer debond lengths compared to bundles with lower VF. This behavior has been demonstrated in studies [38,77,78]. Additionally, if the fibres within the bundle are misaligned, this can further intensify the situation. Misalignment introduces additional stress concentrations at the interfaces between fibres and matrix, which can result in increased debonding lengths and ineffective lengths within the composite bundle. Indeed, the misaligned fibers create discontinuities in the load transfer path, making it easier for cracks to propagate along the interfaces and causing more extensive debonding. Therefore, as it is seen, there is a significant increase in both $l_d$ and IL for MFCD and MFDD bundles compared to AFCD and AFDD bundles.

Based on Fig. 4, for the MFDD bundle, $l_d$ and IL shift both downwards and upwards in comparison with the MFCD bundles. This is attributed to the variation of fiber diameters as it has a significant influence on $l_d$ and IL (SUUP 3). Introducing a distribution of fibre diameters within the composite bundle can help to decrease the effects of fibre misalignment. A variation of fibre diameter can lead to a more gradual transition of stress within the bundle, decreasing localised stress concentrations at the interfaces. This smoother stress distribution can help to prevent the initiation and propagation of debonding, resulting in lower debonding lengths and more effective load transfer within the composite bundle.

Fig.4 illustrates that, fiber diameter distribution and fiber misalignment (AFDD, MFCD, and MFDD bundles) cause significantly higher level of standard deviations of $l_d$ and IL in the bundle. This deviation is even higher for fiber misalignment bundles (MFCD and MFDD).

The average normalized half debond and ineffective lengths ($l_d/2R$, $IL/2R$, with R being fiber radius) of a AFCD bundle reported in [38] are 25 and 85, respectively. In our study, for a similar carbon-epoxy system and applied strain 2%, the average normalized half debond and ineffective lengths of the MFDD bundle are 70 and 110, respectively (see Fig. 4). This demonstrates the influence of both fiber misalignment and variation in fiber diameters.

According to Fig. 3a,c,e, for the MFCD and MFDD bundles, the SCF trends shift downwards in comparison with the AFCD and AFDD bundles. In essence, by increasing debonding and ineffective lengths within a composite bundle (see Fig. 4), the stress concentration factor decreases because the stress is more effectively redistributed and dissipated along the interfaces between the fibres and the matrix, leading to a more uniform stress distribution and reducing the likelihood of localised failure.

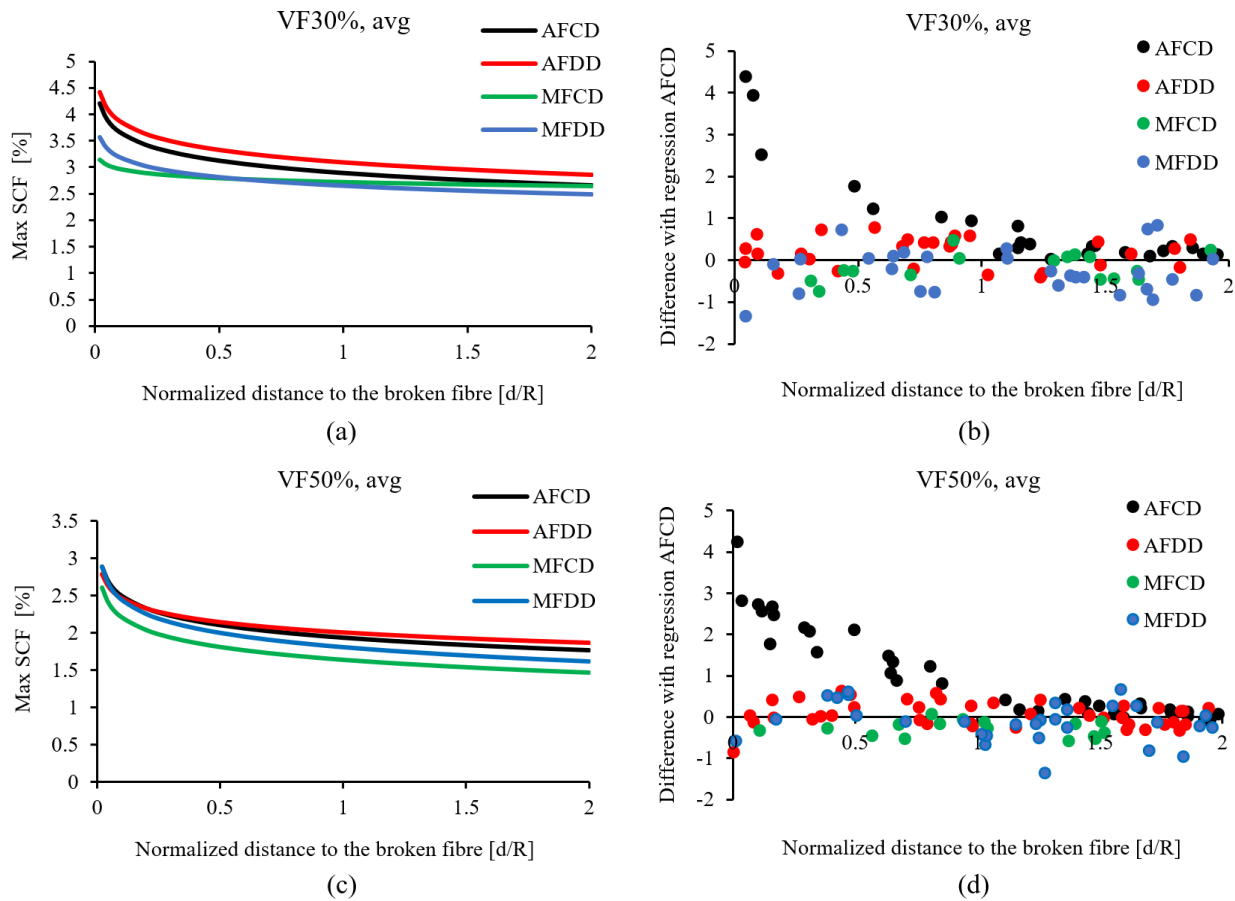

(a) (b) (c) (d)

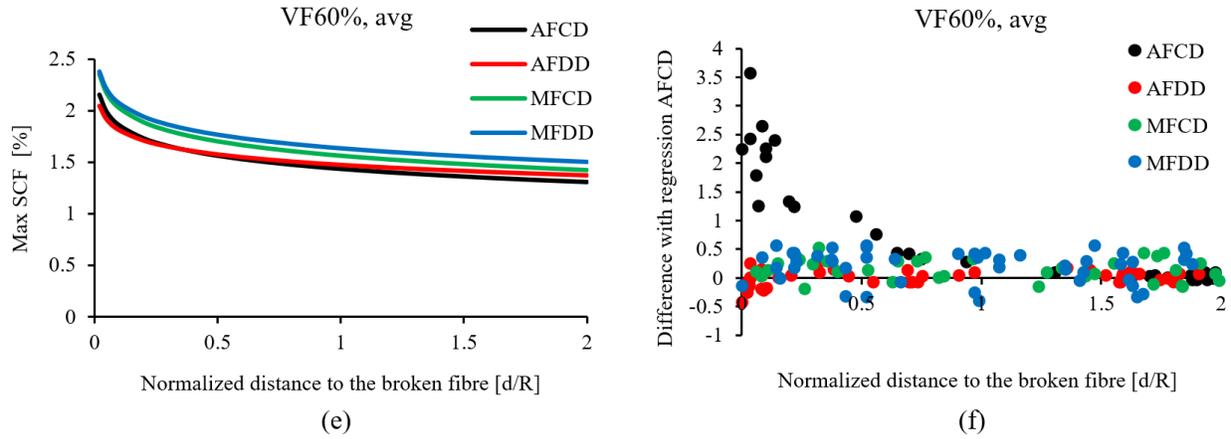

(e)                  (f)

**Fig.3.** Stress concentrations for the AFDD, MFCD, and MFDD bundles compared with the AFCD bundles for different VFs, five random fiber placement realizations for each VF: (a, c, e) SCF regressions in function of the normalized radial distance to the broken fiber and (b, d, f) maximum average SCF ($maxSCF_{avg}$) difference with regression in the AFCD case ($AFDD_{maxSCFs}$, $MFCD_{maxSCFs}$, $MFDD_{maxSCFs}$-ARFCD), each data point represents the difference $AFDD_{maxSCFs}$, $MFCD_{maxSCFs}$, $MFDD_{maxSCFs}$-ARFCD in one intact fiber in one of the realizations of the FE model.

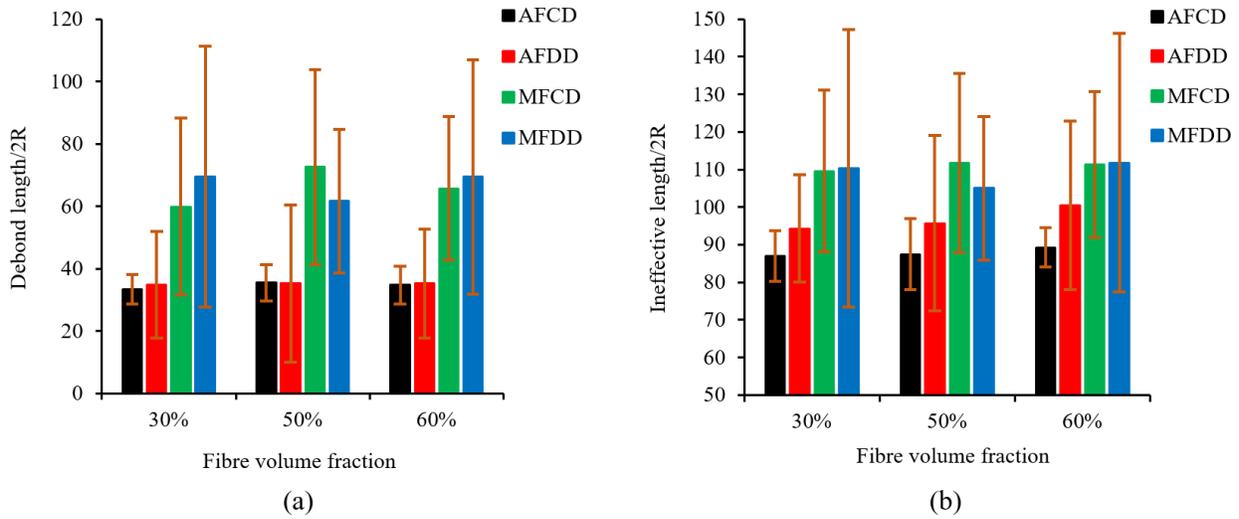

(a)                  (b)

**Fig.4.** Average normalized debond and recovery lengths for the AFDD, MFCD, and MFDD bundles compared with the AFCD bundles, five random fiber placement realizations for each VF, error bars show the standard deviation.

### 3.2. Maximum normal stress in the cross section

As discussed, the reason for the alternative approach of using the maximum stress in a cross-section as a criterion for fiber break, rather than the average stress concentration factor (SCF) taken over the cross section, is based on experimental evidence that supports strong stress localization in composite materials [26]. Using the average stress concentration factor over the cross section as the stress in the neighboring intact fiber may not accurately capture the localized stress concentrations that occur around a broken fiber. By considering the maximum stress in a cross-section as the criterion for fiber break, the fiber break models can better account for the strong stress localization effects observed experimentally. Furthermore, this approach becomes even more crucial when studying fibers with different diameters, as the stress distribution and localization effects can vary significantly based on the size and geometry of the fibers.

In this section, the SCF is calculated based on the peak stress in the cross section, Eq (2), and referred to as $maxSCF_{max}$. Five random realizations of each carbon bundle type at different VFs based on the material properties and FE parameters presented in Table 1 and Table 2 were generated. Fig. 5a,c,e depict SCF trends for the AFDD, MFCD, and MFDD models in comparison with the AFCD models. It is evident that AFDD exhibits a significant decreasing trend, particularly at $d/R < 1$, in contrast to AFCD. However, there is no a clear trend when comparing MFCD and MFDD with the AFCD and AFDD bundles. This is attributed to the location and size of fiber misalignment. Consequently, even a clear trend is not observed among different VFs within the MFCD and MFDD bundles, especially at $d/R < 0.5$.

One notable observation is that the SCF decay is slower in the MFCD and MFDD bundles compared to the AFCD and AFDD bundles, resulting in elevated SCF level in neighboring fibers, which are farther away. Misaligned fibers are not able to efficiently transfer stress to neighboring fibers. When fibers are aligned, they can distribute stress more evenly and effectively, leading to a faster decay of the SCF. However, in misaligned bundles, the stress is not as efficiently transferred, resulting in a slower decay of the SCF and higher stress levels in neighboring fibers that are farther away. This shows that in a bundle with fiber misalignment, a greater number of neighboring fibers are affected by fiber breakage, leading to higher SCF in those fibers. The distribution of SCF alters the local stress redistribution around a fiber break, and may change cluster formation.

Comparing $maxSCF_{max}$ vs $maxSCF_{avg}$ graphs, Fig. 5 vs Fig. 3, it is observed that the peak of the $maxSCF_{max}$ is about twice higher than $maxSCF_{avg}$ at all VFs and bundle types. Moreover, as shown in SUPP 1, the $maxSCF_{max}$ is extremely higher than $maxSCF_{avg}$ in the AFDD perfectly bonded bundle. Both $maxSCF_{avg}$ and $maxSCF_{max}$ when perfectly bonded or with static debonding have been investigated in depth in [78]. These observations suggest that:

i. The primary factor influencing the development of fiber breaks is the maximum stress ($maxSCF_{max}$) within fiber's cross-section, making it crucial to prioritize this criterion over the baseline models [1-3] that rely on $maxSCF_{avg}$ as the maximum stress in neighboring intact fibers.

ii. Debonding significantly decreases the $maxSCF_{max}$ in the NNFs compared to perfectly bonded bundles (refer to [77,78] for more information about perfectly bonded bundles).

iii. Misalignment should be included in the fiber break modeling as it changes the local SCF distribution around fiber breaks.

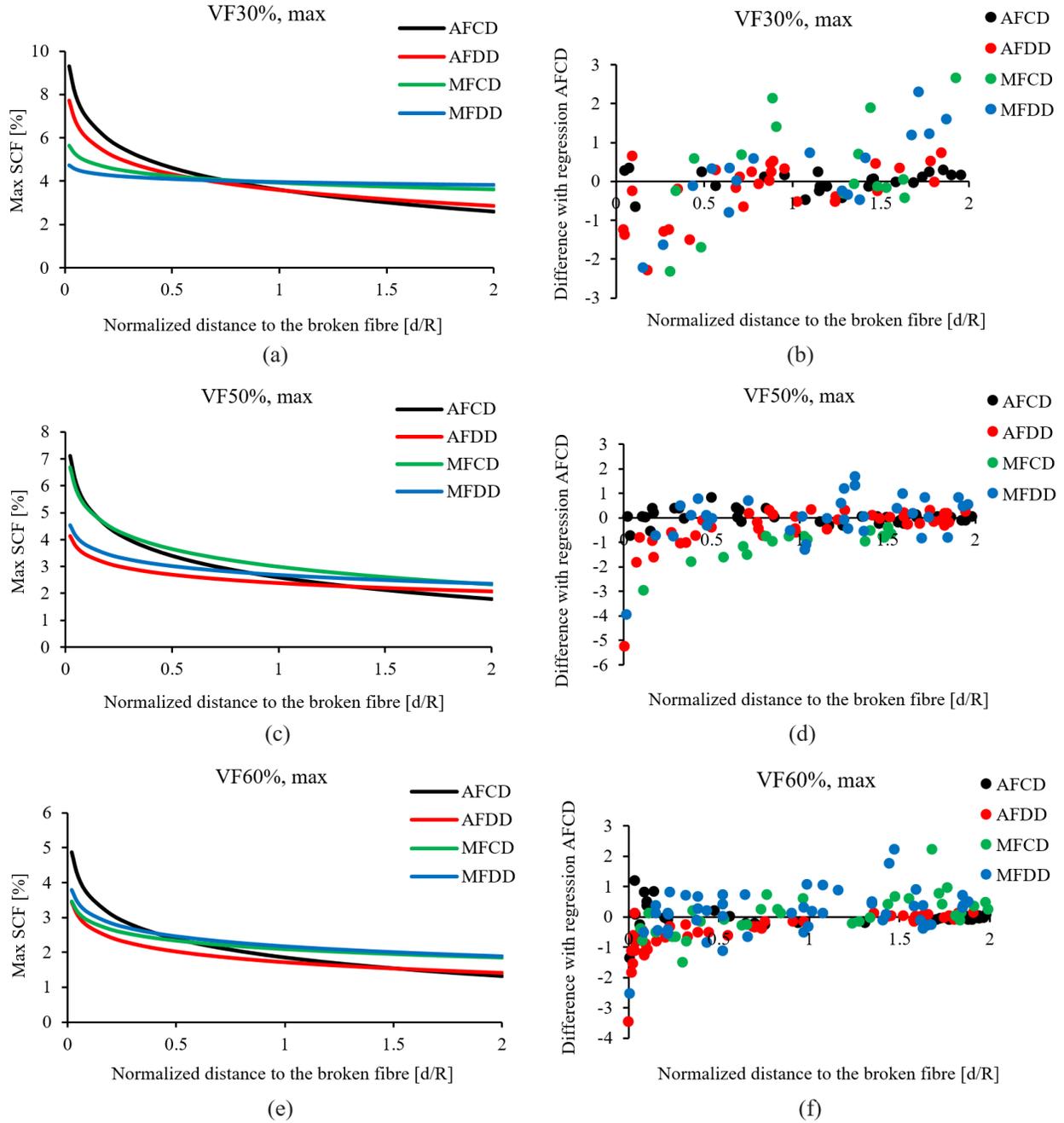

**Fig.5.** Stress concentrations for the AFDD, MFCD, and MFDD bundles compared with the AFCD bundles for different VFs, five random fiber placement realizations for each VF: (a, c, e) SCF regressions in function of the normalized radial distance to the broken fiber and (b, d, f) maximum average SCF ($maxSCF_{max}$) difference with regression in the AFCD case ($AFDD_{maxSCFs}, MFCD_{maxSCFs}, MFDD_{maxSCFs}$-ARFCD), each data point represents the difference $AFDD_{maxSCFs}, MFCD_{maxSCFs}, MFDD_{maxSCFs}$-ARFCD in one intact fiber in one of the realizations of the FE model.

## 4. Statistical analysis of the AFDD, MFCD, and MFDD vs AFCD bundles

A statistical analysis, namely, t-Test is conducted to evaluate the significance of the differences among the AFDD, MFCD, MFDD, and AFCD bundles across the entire trend lines (*d/R*). The comparisons were performed at a confidence level of 95%. The analysis is done based on average and maximum normal stresses ($maxSCF_{avg}$, $maxSCF_{max}$) in the cross section.

Tables. 2 and Table. 3 present a summary of the main findings from the statistical analysis. The tables highlight differences between the calculated SFC's and the regression function of AFCD (RAFCD) bundle, coefficient of variation $CV = \frac{std(SCF_{\frac{\sigma}{2},\sigma,2\sigma} - RFCD)}{\bar{\mu}(SCF_{\frac{\sigma}{2},\sigma,2\sigma} - RFCD)}$ where std and $\bar{\mu}$ denote standard deviation and mean respectively, and p-value for probability of the null hypothesis ($SCF_{AFDD}, SCF_{MFCD}, SCF_{MFDD} = SCF_{FCD}$). The level of confidence for rejection ($1 - p_{value}$) indicates whether there is a statistically significant difference among the AFDD, MFCD, MFDD, and AFCD bundles.

### 4.1. Average normal stress over the cross section

Table. 3 shows the statistical analysis of average normal stress over the cross section. The results indicate a strong rejection based on p-values at a 95% confidence level across the entire range of *d/R* for AFDD, MFCD, MFDD, compared to the AFCD bundles at different VFs.

Table. 4 presents the debond length analysis based on average normal stress over the cross section at different VFs. As it is clear, there is significant increase in $l_d$ for the MFCD and MFDD bundles compared to the AFCD and AFDD bundles at different VFs.

Table. 5 depicts the ineffective length analysis based on average normal stress over the cross section at different VFs. As it is seen, there is significant increase in IL for the MFCD and MFDD bundles compared to the AFCD and AFDD bundles at different VFs.

**Table.3.** The SCF comparison "one tail t-Test with an unequal variance analysis" for $maxSCF_{avg}$ calculations.

| VF | SCF relative difference $\bar{\mu}$ ((AFDD, MFCD, MFDD – RAFCD)/RAFCD) % | | | p-value one tail for the null hypothesis AFCD = AFDD, MFCD, MFDD | | | Is the difference significant at 95% level? | | | CV | | |
|---|---|---|---|---|---|---|---|---|---|---|---|---|
| | AFCD vs AFDD | AFCD vs MFCD | AFCD vs MFDD | AFCD vs AFDD | AFCD vs MFCD | AFCD vs MFDD | AFCD vs AFDD | AFCD vs MFCD | AFCD vs MFDD | AFCD vs AFDD | AFCD vs MFCD | AFCD vs MFDD |
| 30% | 6.35 | -5.96 | -8.83 | 0.007 | 0.0002 | 0.0001 | YES | YES | YES | 1.79 | 1.81 | 2.04 |
| 50% | 2.95 | -15.20 | -6.76 | 6E-06 | 2E-08 | 6E-07 | YES | YES | YES | 5.33 | 0.65 | 3.84 |
| 60% | 0.15 | 8.87 | 13.40 | 3E-05 | 0.0002 | 0.0007 | YES | YES | YES | 21.33 | 1.41 | 1.34 |

**Table. 4.** The $l_d$ comparison between AFDD, MFCD and MFDD bundles and AFCD bundle.

| VF | $\bar{\mu}\,(l_d)$ of 5 random relisations | | | | Std | | | | $l_d$ relative difference $\bar{\mu}\,((\text{AFDD, MFCD, MFDD} - \text{AFCD})/\text{AFCD})\,\%$ | | |
|---|---|---|---|---|---|---|---|---|---|---|---|
| | AFCD | AFDD | MFCD | MFDD | AFCD | AFDD | MFCD | MFDD | AFCD vs AFDD | AFCD vs MFCD | AFCD vs MFDD |
| 30% | 33.4 | 34.9 | 59.9 | 69.6 | 4.73 | 17.07 | 28.35 | 41.84 | 4.6 | 79.6 | 108.6 |
| 50% | 35.5 | 35.3 | 77.6 | 61.7 | 5.79 | 25.21 | 31.23 | 22.98 | -0.5 | 104.6 | 73.9 |
| 60% | 34.8 | 35.2 | 65.7 | 69.4 | 6.11 | 17.38 | 23.05 | 37.58 | 1.2 | 88.8 | 99.7 |

**Table. 5.** The IL comparison among the AFDD, MFCD, and MFDD bundles vs the AFCD bundle, μm.

| VF | $\bar{\mu}\,(\text{IL})$ of 5 random relisations | | | | Std | | | | IL relative difference $\bar{\mu}\,((\text{AFDD, MFCD, MFDD} - \text{AFCD})/\text{AFCD})\,\%$ | | |
|---|---|---|---|---|---|---|---|---|---|---|---|
| | AFCD | AFDD | MFCD | MFDD | AFCD | AFDD | MFCD | MFDD | AFCD vs AFDD | AFCD vs MFCD | AFCD vs MFDD |
| 30% | 87.05 | 94.3 | 109.6 | 110.3 | 6.76 | 14.27 | 21.54 | 36.91 | 8.3 | 25.88 | 26.68 |
| 50% | 87.44 | 95.7 | 111.7 | 105.0 | 9.42 | 23.33 | 23.81 | 19.07 | 9.44 | 27.78 | 20.11 |
| 60% | 89.26 | 100.5 | 111.4 | 111.8 | 5.24 | 22.42 | 19.43 | 34.40 | 12.61 | 24.76 | 25.24 |

### 4.2. Maximum normal stress in the cross section

Table. 6 shows the statistical analysis of maximum normal stress in the cross section. For most of the cases, results do not show a strong rejection based on p-values at a 95% confidence level across the entire range of *d/R* for the AFDD, MFCD, MFDD, compared to the AFCD bundles at different VFs. This is in contrast with the trend line observations presented in section 3.2.

**Table.6.** The SCF comparison "one tail t-Test with an unequal variance analysis" for $maxSCF_{max}$ calculations.

| VF | SCF relative difference $\bar{\mu}\,((\text{AFDD, MFCD, MFDD} - \text{RAFCD})/\text{RAFCD})\,\%$ | | | p-value one tail for the null hypothesis AFCD = AFDD, MFCD, MFDD | | | Is the difference significant at 95% level? | | | CV | | |
|---|---|---|---|---|---|---|---|---|---|---|---|---|
| | AFCD vs AFDD | AFCD vs MFCD | AFCD vs MFDD | AFCD vs AFDD | AFCD vs MFCD | AFCD vs MFDD | AFCD vs AFDD | AFCD vs MFCD | AFCD vs MFDD | AFCD vs AFDD | AFCD vs MFCD | AFCD vs MFDD |
| 30% | -2.45 | 11.47 | 10.36 | 0.07 | 0.2 | 0.27 | NO | NO | NO | 3.34 | 4.56 | 6.38 |
| 50% | -8.94 | -37.36 | 3.93 | 0.0047 | 1.9E-6 | 0.45 | YES | YES | NO | 2.25 | 0.59 | 47.91 |
| 60% | -13.3 | 10.58 | 12.55 | 0.0001 | 0.29 | 0.12 | YES | NO | NO | 1.34 | 8.36 | 4.6 |

## 5. Conclusion

This research investigates the effect of longitudinal debonding between the broken fiber and the matrix on stress redistribution caused by a fiber break in a UD impregnated fiber bundle, taking into account fiber misalignment and fiber diameter distribution in random fiber packings. The research was conducted on carbon-reinforced epoxy bundles according to a finite-element analysis of the stress state of a bundle around a broken fiber in four different bundle types: (i) debonding in aligned fibers with a constant diameter (AFCD) bundle, (ii) debonding in aligned fibers with diameter distribution (AFDD) bundle, (iii) debonding in misaligned fibers with a constant diameter (MFCD) bundle, and (iv) debonding in misaligned fibers with diameter distribution (MFDD) bundle. For each bundle type and each fiber volume fraction value of VF30%, VF50%, and VF60%, five random realizations were built. The stress redistribution was evaluated at strain level of 2% (inelastic regime) based on average and maximum normal stress over the cross section.

Based on average calculations, the AFDD bundles show a higher SCF trend and IL predictions than the AFCD bundles at VF30%, while it diminishes at higher VF50% and VF60%. The MFCD bundles reveals a statistically significant lower SCF trend and higher IL predictions compared to the AFCD bundles at different VFs. Similarly, the MFDD bundles demonstrates a statistically significant lower SCF trend and higher IL predictions in comparison with the AFCD bundles at different VFs, but show an increasing trend when compared as the MFCD vs. AFCD bundles due to the variability in fiber diameters within the bundles, which leads to an increase in stress redistribution parameters.

Based on maximum calculations, all bundle types illustrate a significantly lower SCF trend predictions than AFCD bundles, especially at $d/R < 0.5$. A key observation is that the SCF decay occurs slower in the MFCD and MFDD bundles when compared to the AFCD and AFDD bundles, leading to higher SCF levels in neighboring fibers, which are farther away.

The current FE modeling explores the improvement of the accuracy of simulating the stress redistributions around a broken fiber. This is investigated within a bundle including fiber-matrix debonding, accounting for misaligned fibers and fiber diameter distribution. Our results show different predictions of stress redistribution, which is reported for longitudinal debonding AFCD randomly packed model in [38]. This can help in bridging the discrepancy between numerical and experimental data to predict the longitudinal tensile failure of composites in better alignment with the observable microscale material behavior.